# Topological structure and dynamics of three dimensional active nematics

Guillaume Duclos[1#], Raymond Adkins[2#], Debarghya Banerjee[3,4], Matthew S. E. Peterson[1], Minu Varghese[1], Itamar Kolvin[2], Arvind Baskaran[1], Robert A. Pelcovits[5], Thomas R. Powers[5,6], Aparna Baskaran[1], Federico Toschi[7], Michael F. Hagan[1], Sebastian J. Streichan[2], Vincenzo Vitelli[8], Daniel A. Beller[9*], Zvonimir Dogic[1,2*]

[1]Department of Physics, Brandeis University, 415 South Street, Waltham, Massachusetts 02453, USA.

[2]Department of Physics, University of California, Santa Barbara, California 93111, USA.

[3]Max Planck Institute for Dynamics and Self-Organization, Am Faßberg 17, 37077 Göttingen, Germany.

[4]Instituut-Lorentz, Universiteit Leiden, 2300 RA Leiden, Netherlands.

[5]Department of Physics, Brown University, Providence, Rhode Island 02912, USA.

[6]School of Engineering, Brown University, Providence, Rhode Island 02912, USA.

[7]Department of Applied Physics, Eindhoven University of Technology, 5600 MB Eindhoven, Netherlands.

[8]James Frank Institute and Department of Physics, The University of Chicago, Chicago, Illinois 60637, USA.

[9]Department of Physics, University of California, Merced, California 95343, USA.

[#]These authors contributed equally to the manuscript

*Correspondence to: dbeller@ucmerced.edu, zdogic@ucsb.edu

Point-like motile topological defects control the universal dynamics of diverse two-dimensional active nematics ranging from shaken granular rods to cellular monolayers. A comparable understanding in higher dimensions has yet to emerge. We report the creation of three-dimensional active nematics by dispersing extensile microtubule bundles in a passive colloidal liquid crystal. Light-sheet microscopy reveals the millimeter-scale structure of active nematics with a single bundle resolution and the temporal evolution of the associated nematic director field. The dominant excitations of three-dimensional active nematics are extended charge-neutral disclination loops that undergo complex dynamics and recombination events. These studies introduce a distinct class of non-equilibrium systems whose turbulent-like dynamics arises from the interplay between internally generated active stresses, the chaotic flows and the topological structure of the constituent defects.

Topological defects are universal descriptors of structure formation and dynamics from cosmological to micro scales (*1-5*). They represent discontinuous configurations of a matter field characterized by topological invariants, such as winding numbers. Fluid vortices are examples of such topological defects: the fluid velocity is undefined at the vortex core but it winds by $2\pi$ around any closed loop encompassing the core (*6*). The description of ordered soft materials, such as nematic liquid crystals, necessitates an additional director field that describes the local anisotropy, which enlarges the space of possible defect types. Topological windings of the nematic director field give rise to disclination lines, whose threadlike appearance is the origin of the phase's name. Disclinations, like fluid vortex lines, can form links, knots, and undergo reconnection events (*7-12*).

Equilibrium nematics favor uniform alignment that minimizes distortion energy and leads to annihilation of topological defects. Conversely, in 2D active nematics, the continual injection of energy by propelled anisotropic constituents destabilizes large-scale defect-free uniform alignment (*13*). Instead it generates chaotic flows in which pairs of motile point-like defects are continually being created and annihilated (*8, 14-16*). Such universal dynamics is observed in remarkably diverse 2D systems ranging from millimeter-sized shaken granular rods and micron-sized complex motile living cells to nano-scale motor-driven biological filaments (*17-23*). Several obstacles have hindered progress in generalizing defect-driven dynamics of active nematics to three dimensions. First, the higher dimensionality greatly enriches the manifold of possible defect configurations. Discriminating between different defect types requires measurement of the spatiotemporal evolution of the director field on macroscopic scales, an experimental feat that has not been accomplished even for equilibrium samples. Second, there is a lack of model 3D systems, as many experimental candidates cannot be rendered active away from surfaces. We overcame these limitations by combining the development of a unique microtubule-based active nematic model system with state-of-the-art 3D light sheet microscopy (*24*).

We created a tunable model of 3D active nematics from isotropic active fluids composed of molecular motor-driven microtubule bundles (*17*). Replacing a broadly acting depletant with a specific microtubule crosslinker PRC1 enabled assembly of a uniform composite mixture of low density extensile microtubule bundles (~0.1% volume fraction) and a passive colloidal nematic based on filamentous viruses (Fig. 1A). This strategy is similar to previous work on the living liquid crystal composites of bacterial swimmers and chromonic liquid crystals (*21*). ATP fueled stepping motion of kinesin motors generates microtubule bundle extension and active stresses that drive the chaotic dynamics of the entire system. Birefringence in polarized light microscopy of the system indicates local nematic order (Fig. 1B), in contrast to active fluids lacking the passive liquid crystal component.

Elucidating the spatial structure of 3D active nematics requires imaging of the nematic director field on scales from microns to millimeters. Furthermore, uncovering its dynamics requires acquisition of spatial director maps with high temporal resolution. To overcome these seemingly divergent constraints, we used a multi-view light-sheet microscope (Fig. 1C) (*24*). The spatiotemporal evolution of the nematic director field $\mathbf{n}(\mathbf{x},t)$ was extracted from a stack of fluorescent images using the 3D structure tensor method (SI). Measuring the spatial gradients of the director field identified regions with large elastic distortions (Fig. 1D). Reconstruction of 3D distortion maps revealed that these localized regions mainly formed curvilinear structures, which could either be isolated loops, or belong to a complex network of system-spanning lines (Fig. 1E). As demonstrated later these curvilinear distortions are topological disclination lines characteristic of 3D nematic. Similar structures were observed in numerical simulations of 3D active nematic dynamics, using either a hybrid Lattice Boltzmann method or a finite difference Stokes solver numerical approach (Fig. 1F, SI) (*25*).

Reducing the ATP concentration slowed down the chaotic flows, which revealed the temporal dynamics of the nematic director field. In turn, this identified the basic events governing the dynamics of disclination lines. We focused on characterizing the closed-loop disclinations, as they are the fundamental objects seen to arise or annihilate in the bulk. Isolated loops nucleated and grew from undistorted, uniformly aligned regions (Fig. 2A; Fig. S1A,2A). Likewise, loops also contracted and self-annihilated, leaving behind a uniform region (Fig. 2B, Fig. S1B,2B). Furthermore, expanding loops frequently encountered and subsequently merged with the system-spanning network of distortion lines (Fig. 2C, Fig. S1D,2D). Distortion lines in the network self-intersected and reconnected to emit a new isolated loop (Fig. 2D; Fig. S1C,2C).

Topological constraints require that topological defects can only be created in cancelling sets containing net-zero topological charge. To obey this rule, point-like defects in 2D active nematics always nucleate as pairs of opposite charge (*8*). In 3D, a disclination loop as a whole can either carry a monopole charge or be topologically neutral, depending on its director winding structure. Since charged topological loops can only appear in pairs, nucleation of isolated loops as observed in our system implies that these loops are topologically neutral.

To establish that the curvilinear distortions are nematic disclination loops with no net charge, we characterized their topological structure. In 2D nematics, point-like disclination defects are characterized by *s*, the winding number or a topological charge. The lowest-energy disclinations have $s = \pm 1/2$, which correspond to a $\pi$ rotation of the director field in the same sense or the opposite sense, respectively, as the traversal of any closed path encircling only the defect of interest. In 3D nematics, point-like defects from 2D systems are generalized to disclination lines, affording a broader variety of director configurations. We define ***t*** to be the disclination line's local tangent unit vector. In a plane perpendicular to ***t*** the director field winds by $\pi$ about a direction specified by ***Ω***, the rotation vector, which can make an arbitrary angle *β* with ***t*** (*26*). If ***Ω*** points antiparallel or parallel to ***t***, then the local director field is confined to the plane orthogonal to ***t*** and assumes disclination profiles familiar from 2D nematics. These configurations have local wedge winding character (Fig. 3A). If ***Ω*** is perpendicular to ***t***, the director forms a spatially varying angle away from the orthogonal plane, locally creating a twist winding character (Fig. 3A). Because ***Ω*** may point in any direction relative to ***t***, both ***Ω*** and *β* can vary continuously along a disclination line (Fig. 3A).

For disclination lines forming loops, ***Ω*** can vary continuously providing it returns to its original orientation upon closure, leading to a myriad of possible configurations. A family of loops of particular relevance to 3D active nematics is characterized by a uniform ***Ω*** interpolating between two emblematic geometries: *wedge-twist* and *pure-twist* loops. In the wedge-twist loop, ***Ω*** makes an angle $\gamma = \pi/2$ with the loop normal ***N*** (Fig. 3B). As the disclination's tangent ***t*** rotates by $2\pi$ upon traveling around the loop, the angle *β* made by ***Ω*** with ***t*** varies from 0 (+1/2 wedge) to $\pi/2$ (twist), to $\pi$ (−1/2 wedge), then back to $\pi/2$ and returning to 0 (*26, 27*). The *pure-twist* loop has ***Ω*** uniformly parallel to ***N***, so $\gamma = 0$ and ***Ω*** is perpendicular to ***t*** ($\beta = \pi/2$, twist profile) at all points on the loop (Fig. 3C) (*26, 28*). In this family of loops, the director just outside the loop, $n_{out}$, is also uniform. The lack of a winding of both ***Ω*** and $n_{out}$ implies that the loop is topologically neutral (SI) (*9, 29*).

Experimental measurements of the director field allowed us to fully characterize the topological and geometrical structures of the disclination loops (Fig 4). Analysis of the director field indicated that the distortion lines and loops have the $\pi$ winding indicative of disclinations (Fig. 1E,F), with continuous variation of $\beta$ and of the local winding character. Furthermore, a majority of the analyzed loops were well approximated by the family of curves where $\boldsymbol{\Omega}$ and $\boldsymbol{n}_{out}$ varied little along the loop circumferences. Categorizing loops according to their $\gamma$ values revealed that the entire continuous family from wedge-twist (Fig. 4A, B) to pure-twist (Fig. 4C) was represented, with pure-twist-like loops being slightly more prevalent than the wedge-twist loops (Fig. 4D). As discussed previously, the nucleation of an isolated loop without a topological partner requires its topological neutrality. Structural analysis revealed that this is indeed a case as all 268 experimentally and 94 theoretically analyzed loops carried no charge. This demonstrates that amongst many possible configurations, topologically neutral loops are the dominant excitation mode of 3D active nematics. The same class of loop geometries also dominated the dynamics in our numerical simulations of bulk 3D active nematics and in recently studied confined active nematics (Fig. 4E, F) (*30*).

Nucleation of an isolated topologically neutral loop is the 3D analog of unbinding of a 2D disclination pair. The dynamics results from the same active stress driven self-amplifying bend distortions as in 2D extensile nematics, but the exact mechanisms that give rise to the wedge-twist or pure-twist loop are different. In the wedge-twist loop, a cross-section through the +1/2 and -1/2 wedge profile recalls unbinding of a pair of point-disclinations in 2D (Fig. 5A,B). The +1/2 wedge profile typically appears on the side of the growing bend distortion, oriented away from the -1/2 wedge profile. Similarly, wedge-twist loops with the +1/2 wedge profile oriented inward towards the -1/2 wedge are driven to shrink by active and passive stresses. Unlike in 2D active nematics, after nucleation the wedge profiles remain bound to each other through a disclination loop that includes points with a local twist winding character. Even though local active nematic stresses alone are not expected to drive growth of a pure-twist loop, both simulations and experiments show cases of nearly pure-twist $\gamma \approx 0$ loop creation (Fig 5C,D). We found that chaotic flows built up twist distortions that locally relaxed through creation of a pure-twist loop (Fig 5D,E).

By coupling chaotic flows to an order parameter for local orientational ordering, 3D active nematics represent a new class of non-equilibrium systems that exhibits turbulent-like dynamics.

Combined with other emerging theoretical work (*30, 31*), the developed experimental model system offers a powerful platform to investigate the role of topology, dimensionality, and material order in the chaotic internally driven flows of active soft matter. Furthermore, the use of a multi-view light-sheet imaging technique demonstrates its potential to unravel dynamical processes in diverse non-equilibrium soft materials, such as relaxation of nematic liquid crystals upon a quench or their deformation under external shear flow.

**Acknowledgements:** We thank Bezia Lemma, Radhika Subramanian and Marc Ridilla for their help in protein purification. D.A.B. acknowledges S. Čopar, S. Žumer and M. Ravnik for helpful discussions on disclinations in 3D active nematics. Experimental work was supported by the U.S. Department of Energy, Office of Basic Energy Sciences, through award DE-SC0010432TDD (G.D., R.A. and Z.D.). G.D. also acknowledges support of HFSP long term fellowship. Theoretical modeling was supported by NSF-DMR-1855914, NSF-MRSEC-1420382 and NSF-CBET-1437195 (D.A.B., M.P., M.V. Ar.B, Ap.B, M.F.H, R.P. and T. P.). Computational resources were provided by the NSF through XSEDE computing resources (MCB090163). We acknowledge use of the Brandeis optical, HPCC, and biosynthesis facilities supported by NSF-MRSEC-1420382. DB was supported by FOM and NWO. V.V. was supported by the Complex Dynamics and Systems Program of the Army Research Office under grant W911NF-19-1-0268. S.J.S. was supported by a NIH-R00 award (5R00HD088708-05). D.A.B. thanks the Isaac Newton Institute for Mathematical Sciences for support during the program "The mathematical design of new materials," supported by EPSRC grant number EP/R014604/1. G.D., I.K., S.J.S. and R.A. conducted experimental research; together with M.S.P. and D.A.B. they analyzed experimental and theoretical data. D.B., F.T. and V.V. developed Lattice Boltzmann simulations. M.S.P., M.V., Ar.B, Ap.B and M.F.H. developed and applied finite difference Stokes solver code. D.A.B., R.P. and T.R.P. conducted theoretical modeling and interpretation of data. G.D., V.V., Ap.B., M.F.H., D.A.B. and Z.D conceived the work. G.D., D.A.B., V.V. and Z.D. wrote the manuscript. All authors have reviewed the manuscript.

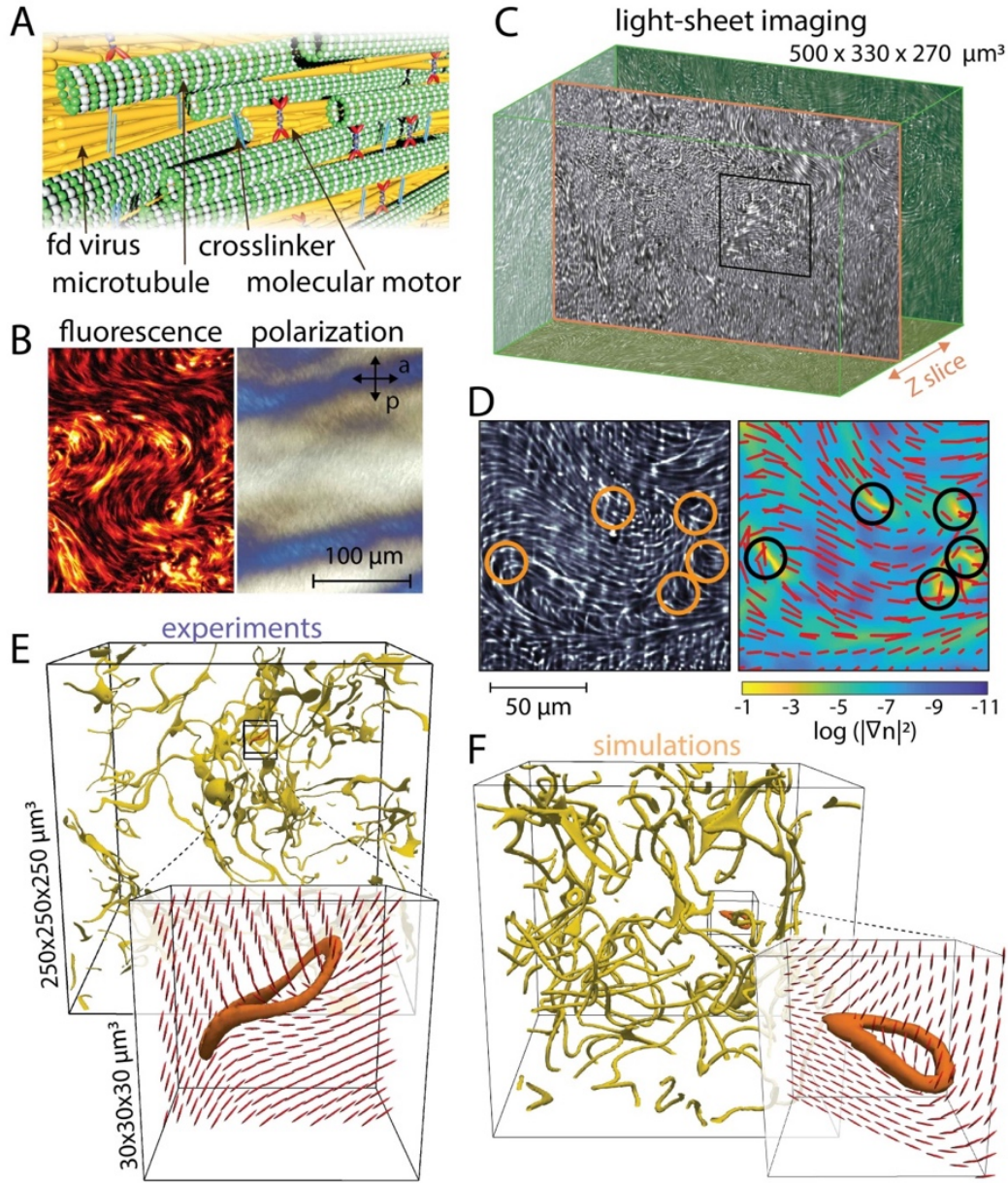

**Fig. 1:** Assembling 3D active nematics and imaging the director field. (**A**) Schematic of the 3D active nematic system: stress generating extensile microtubule bundles are dispersed in a passive colloidal liquid crystal. (**B**) Active nematic imaged with widefield fluorescent microscopy (left) and polarized microscopy (right). Birefringence indicates local nematic order. (**C**) Multi-view light sheet microscopy allows for 3D imaging of millimeter sized samples with single-bundle resolution. (**D**) (left) A 2D slice of fluorescent microtubule bundles with highlighted local elastic distortions. (right) The elastic distortion energy map, with an overlaid nematic director field (red). (**E**) Three-dimensional elastic distortion map suggests the presence of curvilinear rather than point-like singularities. An entangled network of lines coexists with isolated loops. (**F**) Lattice Boltzmann simulations yield a similar structure of 3D active nematics.

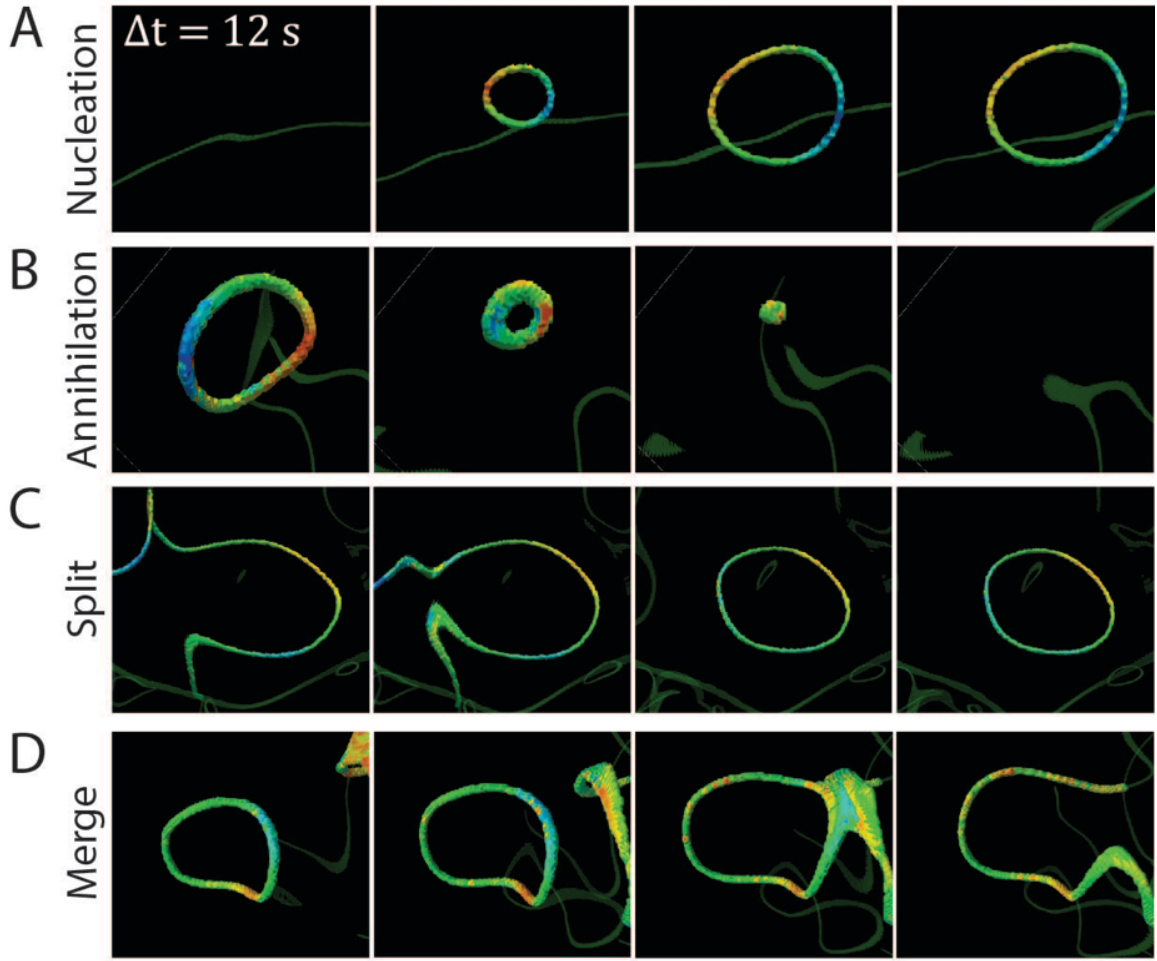

**Fig. 2:** Dynamics of experimentally observed disclination loops. (**A**) Loop nucleation from a defect-free region. (**B**) Loop self-annihilation leaves behind a defect-free nematic. (**C**) A disclination line self-intersects, reconnects, and emits a loop. (**D**) A disclination loop intersects, reconnects and merges with a disclination line. Coloring indicates the angle $\beta$ defined in Fig. 3. Scales and bounding boxes are shown in Fig. S3.

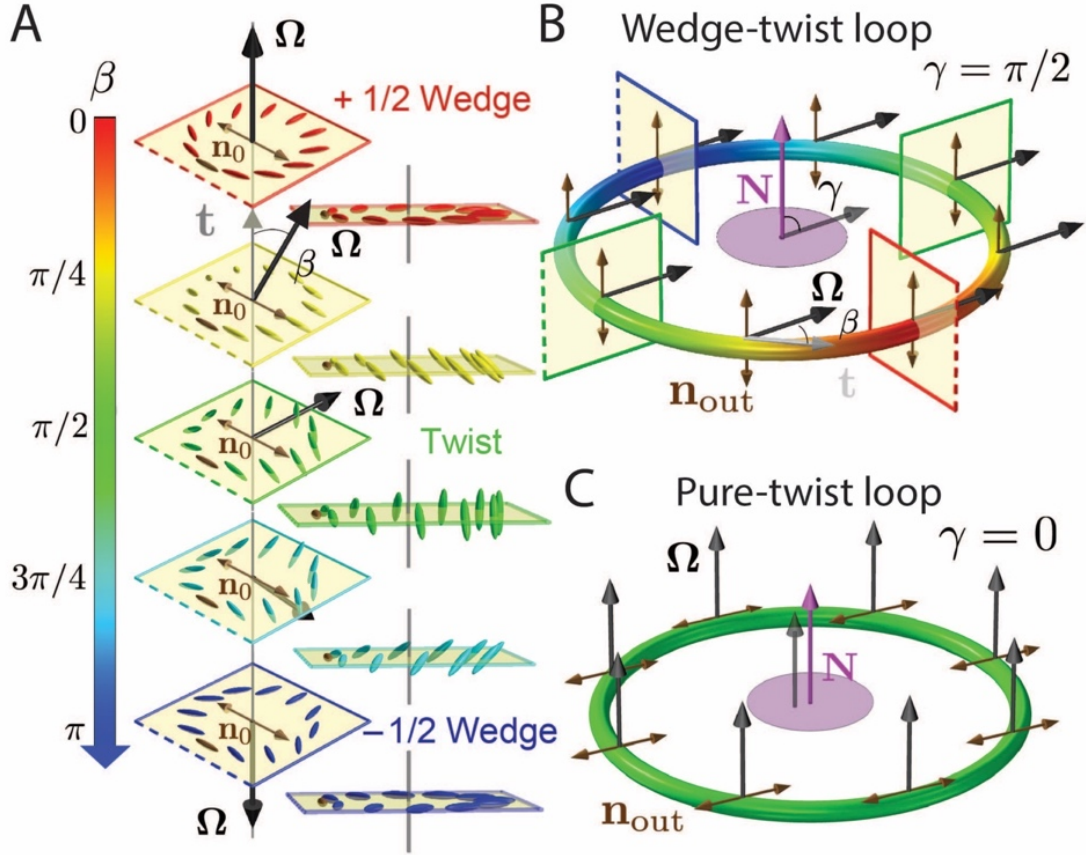

**Fig. 3:** Structure of disclinations lines, wedge-twist and pure-twist loops. (**A**) Disclination line where a local +1/2 wedge winding character continuously transforms into -1/2 wedge through an intermediate twist winding character. The director field winds by $\pi$ about the rotation vector $\mathbf{\Omega}$ (black arrows), which makes angle $\beta$ with the tangent $\mathbf{t}$ (grey arrow). For ±1/2 wedge windings, $\beta=0$ and $\pi$. $\beta=\pi/2$ indicates local twist winding. Reference director $\mathbf{n}_0$ (brown) is held fixed. Color map indicates angle $\beta$. (**B**) The wedge-twist loop where angle $\beta$ varies along the loop. $\mathbf{\Omega}$ is spatially uniform and forms an angle $\gamma=\pi/2$ with the loop's normal, $N$. The winding character in the four illustrated planes corresponds to the profiles of the same colors shown in (A), with dashed edges of squares aligned to show the local director field. Double-headed brown arrows indicate $\mathbf{n}_{out}$, the director just outside the loop. (**C**) Pure-twist loop, with $\mathbf{\Omega}$ both uniformly parallel to loop normal $N$ ($\gamma=0$) and perpendicular to the tangent vector.

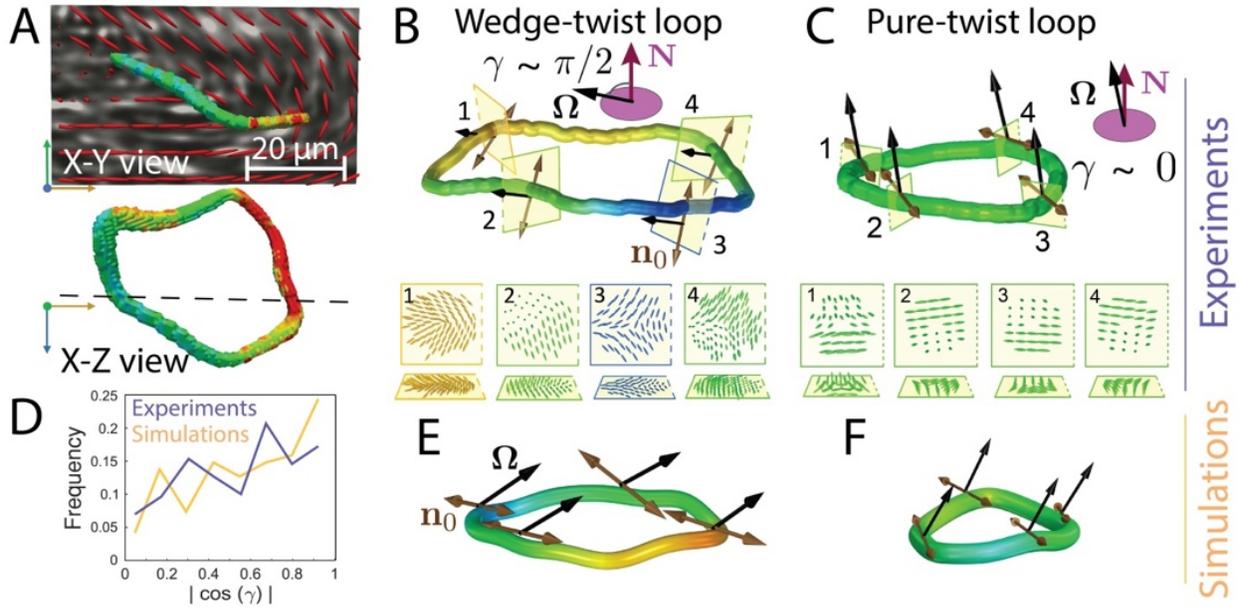

**Fig. 4:** Structure of disclination loops in experiments and theory. (**A**) Two orthogonal views of an experimental wedge-twist loop overlaid onto a fluorescent image of the microtubules. The extracted nematic director is in red. (**B** and **E**) Structure of wedge-twist type disclination loops in experiments and simulation. (**C** and **F**) Structure of pure-twist type disclination loops from experiment and simulation Panels show the director field's winding character in the corresponding cross-sections on the loops of B and C. (**D**) Distribution of loop types extracted from experiment (N=268) and simulation (N=94). $|\cos(\gamma)|=0$ for wedge-twist loops and 1 for pure-twist loops. Distributions of standard deviations of $|\cos(\gamma)|$ are shown in Fig. S4. Coloring indicates the angle $\beta$ as in Fig. 3.

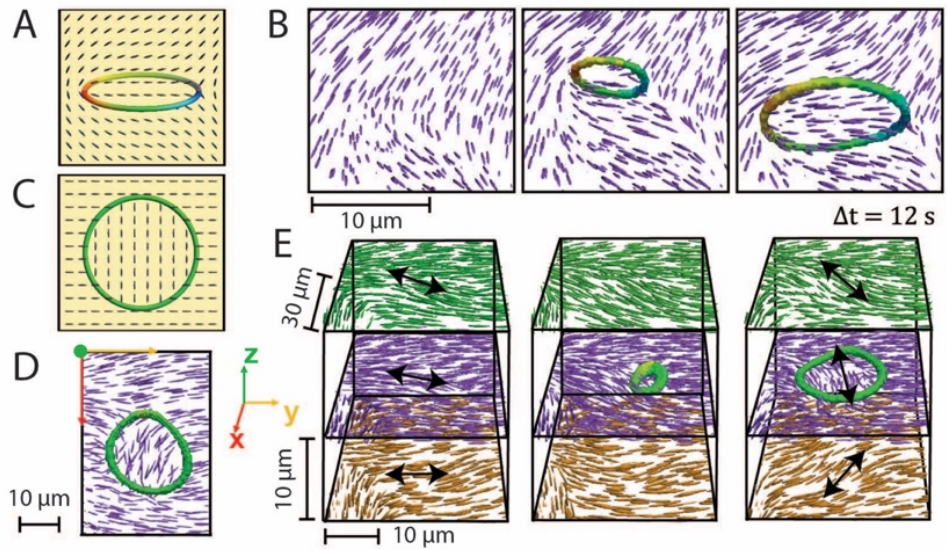

**Fig. 5:** Nucleation mechanism of wedge-twist and pure-twist loops in experiments. (**A**) Schematic of a wedge-twist loop and the director field in the plane that intersects ±1/2 wedge profiles. (**B**) Nucleation and growth of a wedge-twist disclination loop in a bend-distorted region. Purple rods represent the 2D director field through the local ±1/2 wedge profiles. (**C**) Schematic of a pure-twist loop and the director field in the loop's plane. (**D**) Top view of a growing twist disclination loop (same as third panel in E). (**E**) A pure-twist disclination loop nucleates and grows from a local twist distortion. Black arrows indicate the buildup of twist distortions.

# Supplementary Information
# Topological structure and dynamics of three dimensional active nematics

Guillaume Duclos[1,#], Raymond Adkins[2,#], Debarghya Banerjee[3,4], Matthew S. E. Peterson[1], Minu Varghese[1], Itamar Kolvin[2], Arvind Baskaran[1], Robert A. Pelcovits[5], Thomas R. Powers[5,6], Aparna Baskaran[1], Federico Toschi[7], Michael F. Hagan[1], Sebastian J. Streichan[2], Vincenzo Vitelli[8], Daniel A. Beller[9,*], and Zvonimir Dogic[1,2,*]

[1] *Department of Physics, Brandeis University, 415 South Street, Waltham, Massachusetts 02453, USA*
[2] *Department of Physics, University of California, Santa Barbara, California 93111, USA*
[3] *Max Planck Institute for Dynamics and Self-Organization, Am Faßberg 17, 37077 Göttingen, Germany*
[4] *Instituut-Lorentz, Universiteit Leiden, 2300 RA Leiden, Netherlands*
[5] *Department of Physics, Brown University, Providence, Rhode Island 02912, USA*
[6] *School of Engineering, Brown University, Providence, Rhode Island 02912, USA*
[7] *Department of Applied Physics, Eindhoven University of Technology, 5600 MB Eindhoven, The Netherlands*
[8] *James Frank Institute and Department of Physics, The University of Chicago, Chicago, Illinois 60637, USA*
[9] *Department of Physics, University of California, Merced, California 95343, USA*
[#] *These authors contributed equally to the manuscript*
[*] *Correspondence to: dbeller@ucmerced.edu, zdogic@ucsb.edu*

## 1 Experimental methods

**Microtubule polymerization:** Tubulin dimers were purified from bovine brains, through two cycles of polymerizations and depolymerizations in a high molarity PIPES buffer (*1*). Tubulin was flash frozen and stored at -80 °C in M2B buffer (80 mM PIPES, pH 6.8, 1 mM EGTA, 2 mM $MgCl_2$). Tubulin was labeled with Alexa-Fluor 647-NHS (Invitrogen, A-20006) as previously described (*2*). To induce polymerization, tubulin (80uM in M2B) was mixed with 0.6mM GMCPP (Jena Biosciences, NU-4056) and 1mM dithiothreitol DTT in M2B. 3% of the tubulin was fluorescently labeled. The tubulin was first incubated for 30 min at 37°C, followed by an annealing step at room temperature for 6 hours. The GMPCPP stabilized microtubules have an average length of 1.5 $\mu$m (*3*). The polymerized microtubules were then flash frozen and stored at -80°C.

**Assembly of kinesin clusters:** K401-BIO-6xHIS is the 401 amino acid N-terminal domain derived from the Drosophila melanogaster kinesin-1 and labeled with a 6-his and a biotin tag. The motor proteins were transformed, expressed in Rosetta (DE3) pLysS *E. coli* cells, and purified as described previously (*4*). The purified proteins were flash frozen in liquid nitrogen with 36% sucrose and stored at -80 °C. The K401 biotin-labeled motors were assembled into multimotor clusters using tetravalent streptavidin (ThermoFisher, 21122). We mixed 5 $\mu$L of 6.4 $\mu$M kinesin motors with 5.7 $\mu$L of 6.6 $\mu$M streptavidin (MW: 52.8 kDa) in 1.7:1 biotin to streptavidin ratio, and 0.5 $\mu$L of 5 mM dithiothreitol (DTT) in M2B (80 mM PIPES,pH 6.8, 1 mM EGTA, 2 mM $MgCl_2$). The clusters were then incubated on ice for 30 min, before being flash frozen and stored at -80C.

**PRC-1 purification:** The truncated PRC1-NS$\Delta$C (MW: 58 kDa) was used to specifically crosslink mi-crotubules while still allowing for their sliding. It consists of the first 486 amino acids of the full length PRC1 protein (MW: 72.5 kDa). The protein was transformed and expressed in Rosetta BL21(DE3) cells, and subsequently purified as described previously (*5*). The truncated form of PRC1 has greatly increased stability, while maintaining the dimerization and the microtubule binding domain.

**Colloidal virus purification:** Filamentous viruses were grown in bacteria and purified as described previously (*6*). All samples were stored in high salt 20 mM Tris buffer (pH 8.0, 100 mM NaCl). The virus concentration was determined using absorption spectroscopy.

**Assembling active nematic liquid crystal:** We assembled the active nematic liquid crystal by doping an *fd* colloidal liquid crystal with extensile microtubule bundles. A pre-mixture was prepared in a high salt M2B (M2B + 3.9 mM $MgCl_2$) containing an oxygen scavenging system (3.3 mg/mL glucose, DTT (5.5 mM), glucose oxidase (Sigma, G2133) and catalase (0.038 mg/mL, Sigma, C40)), 2nM Trolox to reduce photobleaching (Sigma, 238813), ATP (1420 uM), an ATP regeneration system (phosphoenol pyruvate (26 mM PEP, Beantown Chemical, 129745) and pyruvate kinase/lactate dehydrogenase enzymes (2.8% v/v PK/LDH, Sigma, P-0294)). We then added the kinesin clusters (121 nM Streptavidin), PRC-1 (200nM), *fd* viruses (25 mg/mL) and microtubules (13 $\mu$M). Typical samples consist of 7 mL of Premix, 3 mL of fd virus (100 mg/mL) and 2mL of microtubules (8 mg/mL). The samples are typically active for 3 hours. Passive samples were assembled using the same recipe but lacked both ATP and the kinesin clusters. The isotropic samples were obtained by assembling the previous recipe without fd-wt viruses. For SPIM imaging, the ATP concentration was reduced to 100 $\mu$M. The acquisition time (12sec) of a 3D stack was small compared to the typical fluid velocity (less than 0.1 $\mu$m/sec).

**Widefield microscopy:** Samples were imaged with an inverted microscope (Nikon Ti-E) equipped with a XYZ motorized stage, polarization optics, and a fluorescence imaging module. Simultaneous pictures of the birefringence and the fluorescent microtubules were obtained with a 20x objective (Pan Fluor, NA 0.75) and a CCD camera (Andor, Clara E).

**Multiview-SPIM imaging:** The 3D samples were imaged using multi-view light sheet microscopy (*7*). Briefly, the microscope is composed of two detection and illumination arms. The detection arm is an epifluorescence microscope, consisting of a water-dipping objective (Apo LWD 25x, NA 1.1, Nikon Instruments Inc.), a filter wheel (HS-1032, Finger Lakes Instrumentation LLC), emission filters (BLP01-488R-25, BLP02-561R-25, Semrock Inc.), a tube lens (200 mm, Nikon Instruments Inc.), and an sCMOS camera (ORCA-Flash4.0 V3 Digital) with an effective pixel size of 0.26 $\mu$m. The illumination arm consisted of a water-dipping objective (CFI Plan Fluor 10x, NA 0.3), a tube lens (200 mm, both Nikon Instruments Inc.), a scan lens (S4LFT0061/065, Sill optics GmbH and Co. KG), a galvanometric scanner (6215 hr, Cambridge Technology Inc.), and two lasers (06-MLD 488 nm, Cobolt AB, and 660LX/LS OBIS 660 nm, Coherent Inc.). The FEP tubing is translated using a linear piezo stage (P-629.1cd with E-753 controller) and rotated using a rotational piezo stage (U-628.03 with C-867 controller) and a linear actuator (M-231.17 with C-863 controller, all Physik Instrumente GmbH and Co. KG). Samples were recorded using 4 views, by 900 rotated views, 200 slices with an optical sectioning of 2 $\mu$m, and a temporal resolution of either 12 or 20s depending if tracer beads were also imaged.

**Image post-processing: data fusion and deconvolution:** Image processing was performed using Matlab, Fiji and associated multi-view deconvolution plugin (*8, 9*). The light sheet data was unpacked and binned (2*2 binning, spatial resolution 0f 0.52 $\mu$m after binning) using a custom made Matlab program. To ensure a correct measurement of the microtubule orientation, we first removed distortions introduced by an anisotropic point-spread function (PSF) - SPIM being subject to optical aberration. We deconvolved the four views post-acquisition using an empirical PSF measured on passive fluorescent particles immersed in the active LC (20 iterations). Finally, the same markers were used to register the four complementary views of sample taken at 90 deg rotation intervals, which also reduced the effect of the anisotropic PSF.

**3D Orientation analysis:** The 3D nematic director field was extracted by computing the local structure tensor in 3D with a custom-made Matlab code. This method has been described previously for 2D samples (*10*). The Gaussian window size was 6um. Choosing the right window size is important to correctly detect the topological defects. Choosing a smaller window size resulted in false positives. Choosing a larger window size led to smoothing of the nematic director, and increased the number of undetected defects. To find the defects, we first computed the distortion energy in the one-elastic constant approximation, approximating spatial partial derivatives $\partial_i$ with finite differences. We did not measure $(\partial_i n_j)(\partial_i n_j)$. Instead we measured $(\partial_i [n_j n_k])(\partial_i [n_j n_k])$. In principle these two provide the same information. However, the first option picks up a large artificial derivative when n flips to -n, and then squares that derivative. The second option fixes the n = -n problem by taking derivatives of $n_j n_k$ instead of just $n_j$, removing the large artificial derivatives. We then looked at the winding of the director field where the distortion energy was greater than a 0.5 threshold. We verified that the defects' detection did not depend on the energy treshold chosen. We manually verified that the defects observed in the nematic field correspond to disclinations in the fluorescent microtubule channel. We found that about 20% of the detected defects were false positives, principally due to a low signal/noise ratio in regions where the light sheet is not in focus (on the edge of the FOV).

## 2   Numerical methods

**Theoretical model:** The state of an active nematic is defined by a flow velocity field **u** and a tensor order parameter field $Q_{ij}$ that gives the orientation of the nematics in the flow. The evolution of these two fields is modeled with the Beris-Edwards equations with an added active stress term: the flow field evolves according to the Navier-Stokes equation with additional stress terms due to the nematic order and activity. Meanwhile, $Q_{ij}$ evolves in the flow by advection and free energy relaxation. We can write the equations for active nematics (*11*) as:

$$\left(\partial_t + u_k \partial_k\right) Q_{ij} - S_{ij} = \Gamma H_{ij}, \tag{1}$$
$$\rho \left(\partial_t + u_k \partial_k\right) u_i = \partial_j \Pi_{ij}. \tag{2}$$
$$\partial_i u_i = 0. \tag{3}$$

Here, $\rho$ is the density (assumed constant), $\Gamma$ is a rotational diffusion constant, $\Pi_{ij}$ is the total stress tensor, and $S_{ij}$ is the generalized advection term:

$$S_{ij} = (\xi E_{ik} + \Omega_{ik})(Q_{kj} + \delta_{kj}/3) + (Q_{ik} + \delta_{ik}/3)(\xi E_{kj} - \Omega_{kj}) - 2\xi (Q_{ij} + \delta_{ij}/3)(Q_{kl} \partial_k u_l). \tag{4}$$

In this expression, $\xi$ is a constant determining flow-aligning or tumbling behavior, and the symmetric and the anti-symmetric spatial derivatives of the velocity field are:

$$E_{ij} = \frac{1}{2}\left(\partial_j u_i + \partial_i u_j\right),$$
$$\Omega_{ij} = \frac{1}{2}\left(\partial_j u_i - \partial_i u_j\right).$$

The relaxation due to the free energy is given by the molecular field $H_{ij}$, which is defined as:

$$H_{ij} = -\delta \mathcal{F}/\delta Q_{ij} + (\delta_{ij}/3)\text{Tr}(\delta \mathcal{F}/\delta Q_{kl}), \tag{5}$$
$$\mathcal{F} = K(\partial_k Q_{ij})^2/2 + A Q_{ij} Q_{ji}/2 + B Q_{ij} Q_{jk} Q_{ki}/3 + C(Q_{ij} Q_{ji})^2/4. \tag{6}$$

In the free energy, $K$ is the single Frank elastic constant, and the Landau expansion coefficients $A, B, C$ control the isotropic-nematic phase transition.

The stress tensor $\Pi_{ij}$ consists of a sum of the viscous stress, the passive stress, and the active stress. The active stress is given by $\Pi_{ij}^a = -\zeta Q_{ij}$, where the activity parameter $\zeta$ is positive for extensile systems and negative for contractile systems. The passive stress is

$$\begin{aligned}
\Pi_{ij}^p &= -p\delta_{ij} + 2\xi(Q_{ij} + \delta_{ij}/3)(Q_{kl} H_{lk}) \\
&\quad -\xi H_{ik}(Q_{kj} + \delta_{kj}/3) - \xi(Q_{ik} + \delta_{ik}/3) H_{kj} \\
&\quad -\partial_i Q_{kl}(\delta \mathcal{F}/\delta \partial_j Q_{lk}) + Q_{ik} H_{kj} - H_{ik} Q_{kj},
\end{aligned}$$

The viscous stress given by $\Pi_{ij}^{\text{vis}} = 2\mu E_{ij}$.

**Hybrid lattice Boltzmann method:** We solve the above set of equations using a hybrid lattice Boltzmann method in the limit of high viscosity (to ensure low Reynolds number) (*11*). The hybrid lattice Boltzmann method employs a lattice Boltzmann method to evolve **u** according to the (modified) Navier-Stokes equation and a finite difference method to calculate the evolution equation of the $Q$–tensor. The feedback of the $Q$–tensor field enters the Navier-Stokes equation as a forcing term. The time integration of the $Q$–tensor field is done using a second order Adams-Bashforth method. The parameters used for the simulations are $\Gamma = 0.33$, $\mu = 0.67$, $\xi = 0.9$, $K = 0.005$, $\zeta = 0.01$, $-B = C = 0.1$, and $A = 0$. Incompressibility ($\partial_j u_j = 0$) of the flow is ensured by keeping the Mach number sufficiently low.

**Zero Reynolds number limit: Finite difference Stokes solver method:** The phenomenology observed in the experiments described here can be predicted from a minimal model for active nematics where the flows are quasi-static and purely due to viscous and active stresses. To demonstrate this, we consider Equation 1 with terms up to second order in the dynamic variables, coupled to an incompressible Stokes flow driven entirely by viscous and active stresses. Non-dimensionalizing the equations gives the following dimensionless parameters: $K^* = \Gamma t_0 K/l_0^2$, $\mu^* = \frac{\mu t_0 l_0}{m_0}$, $\zeta^* = \frac{\zeta l_0 t_0^2}{m_0}$ $A^* = \Gamma t_0 A$, $B^* = \Gamma t_0 B$, $C^* = \Gamma t_0 C$. We chose $K^* = 1$, $\mu^* = 1$, $\zeta^* = 0.2$ and $A^* = -1/3, B^* = -4, C^* = 4$ so that our simulation units are given by $t_0 = 4/(\Gamma C)$, $l_0 = 2\sqrt{K/C}$, $m_0 = 8\mu\sqrt{K}/(\Gamma\sqrt{C^3})$. To ensure numerical stability while solving Equation 1, we use a semi-implicit finite difference time stepping scheme based on a convex splitting of the nematic free energy (*12*). To solve the Stokes equation and enforce incompressibility, we implement a Vanka type box smoothing algorithm on a staggered grid (*13*). The solution at each time step is found using Gauss Seidel relaxation iterations, and the rate of convergence to the solution is accelerated by our multigrid solver.

## 3 Theoretical calculations

**Rotation vector:** The rotation vector $\boldsymbol{\Omega}$ used in this work is the rotation angle specifying the symmetry operation in order parameter space, $\mathbb{R}P^2$, undergone by the director $\mathbf{n}(\mathbf{x})$, on a small closed measuring circuit $C$ around the disclination at the point of interest (*14*). For all disclinations studied in this work, the angle of rotation is $|\boldsymbol{\Omega}| = \pi$, which corresponds to topological winding numbers of $\pm 1/2$. The axis of rotation, $\hat{\boldsymbol{\Omega}} \equiv \boldsymbol{\Omega}/|\boldsymbol{\Omega}|$, is determined from the closed loop $\Gamma$ in $\mathbb{R}P^2$ traced out by $\mathbf{n}$ on the measuring circuit $C$. Because $\Gamma$ is, in general, approximately a semi-great circle, $\hat{\boldsymbol{\Omega}}$ is simply the normal to the semi-circle, with sign given by the positive sense of rotation on $\Gamma$ as $C$ is traversed in the positive sense (Fig. 2D). Near the defect point of interest, $\mathbf{n}$ is (approximately) confined to the plane orthogonal to $\boldsymbol{\Omega}$. We can form an orthonormal triad $\{\mathbf{n}_0, \mathbf{n}_1, \hat{\boldsymbol{\Omega}}\}$ such that, along $C$, $\mathbf{n}$ rotates from $\mathbf{n}_0$ through $\mathbf{n}_1$ into $-\mathbf{n}_0$.

**Topological classification of disclination loops:** The classification of topological defects in this work follows the standard application of homotopy theory to nematics. Disclination loops, like hedgehog point defects, carry an integer hedgehog charge:

$$d = \frac{1}{4\pi} \int_{\mathbb{S}^2} d\theta d\phi \, \mathbf{n} \cdot [\partial_\theta \mathbf{n} \times \partial_\phi \mathbf{n}], \tag{7}$$

identifying the defect with an element of the second homotopy group $\pi_2(\mathbb{R}P^2) \cong \mathbb{Z}$ (*15*). Whereas hedgehogs with different $d$ are topologically distinct, for disclination loops the topological categorization depends only on $d$ modulo 2 (*15*). We refer to loops with even $d$ as topologically neutral, because shrinking a $d = 0$ loop to a point leaves behind a locally defect-free director field.

The topological information about a disclination loop is obtained using a meauring torus $\mathbb{T}^2$ enclosing the defect. Maps from $\mathbb{R}P^2$ to $\mathbb{T}^2$ are divided into four topologically distinct classes with a $\mathbb{Z}_4$ group structure when joining loops together (*16*). We classify a disclination loop's topology by recording $\boldsymbol{\Omega}$ and $\mathbf{n}_{\text{out}}$, which together specify the disclination profile, at several sampled points along the disclination's length. The choice of $\mathbf{n}_{\text{out}}$ as reference director follows the canonical choice of tracking $\mathbf{n}$ along a cycle of the torus unlinked with the disclination (*16*). Together, $\{\mathbf{n}_{\text{out}}, \mathbf{n}_{\text{in}}, \hat{\boldsymbol{\Omega}}\}$ define an orthonormal frame $\mathbf{F}$, where $\mathbf{n}_{\text{in}} \equiv \hat{\boldsymbol{\Omega}} \times \mathbf{n}_{\text{out}}$ is the director on the inner side of the loop. Importantly, $\tilde{\mathbf{F}} \equiv \{-\mathbf{n}_{\text{out}}, -\mathbf{n}_{\text{in}}, \hat{\boldsymbol{\Omega}}\}$ gives the same profile as $\mathbf{F}$ because of the $\mathbf{n} \equiv -\mathbf{n}$ symmetry of the director. From point to point on a disclination loop parametrized by angle $\theta$, the frame $\mathbf{F}$ rotates according to $\mathbf{F}(\theta_{i+1}) = \mathbf{R}_{i,i+1}\mathbf{F}(\theta_i)$, where $\mathbf{R}$ is an element of the group $SO(3)$ of rotations of a 3D rigid body. Upon traversing the loop completely, $\mathbf{F}(\theta = 2\pi)$ must return to either $\mathbf{F}(\theta = 0)$ or $\tilde{\mathbf{F}}(\theta = 0)$. In the latter case, the disclination loop is linked by another disclination loop (or an odd number of them), a situation that we do not observe in experimental or simulated active nematics. If $\mathbf{F}$ returns to itself, then the loop is unlinked, and falls into the even-$d$ or the odd-$d$ class.

Let $\Upsilon$ be the path in $SO(3)$ representing the composition of all the rotations between $N$ sampled points, $\mathbf{R}_{0,2\pi} = \mathbf{R}_{0,\theta_1}\mathbf{R}_{\theta_1,\theta_2}\cdots\mathbf{R}_{\theta_{N-1},2\pi}$. To topologically classify these paths, we lift $\Upsilon$ from $SO(3)$ to the simply connected covering space $SU(2)$, which can be parametrized by the unit quaternions ($\pm i$, $\pm j$, $\pm k$, $\pm 1$, and their multiplicative products). Following the spirit of Ref. (*17*), we choose a parametrization in which $\pm k$ represent the rotation of $\mathbf{n}_{\text{out}}$ by $\pi$ about $\pm\boldsymbol{\Omega}$. The set of all possible total rotations $\mathbf{R}_{0,2\pi}$ is then represented by $\{1, k, -1, -k\} = k^\nu$, $\nu \in \{0, 1, 2, 3\}$ (*16*, *17*), forming a group under multiplication with the requisite $\mathbb{Z}_4$ structure. The mod-4 integer $\nu$ provides a topological index for the loop: topologically neutral loops have $\nu = 0$, topologically charged loops have $\nu = 2$, and the linked-loop scenario corresponds to $\nu = 1$ or $3$. Disclination loops with approximately uniform $\boldsymbol{\Omega}$ and $\mathbf{n}_{\text{out}}$, which are the focus of this work, are trivially $\nu = 0$ loops, and therefore neutral and unlinked. The practical application of this theory, to calculate $\nu$ in observed disclination loops, is described below in Sec. 4.

**A local formula for $\beta$, the angle between rotation vector $\boldsymbol{\Omega}$ and unit tangent t:** Experimental and simulated defect sets presented in this work are color-coded by $\beta$. The tangent vector is calculated straightforwardly from the separation between neighboring points on the disclination contour. We compute $\boldsymbol{\Omega}$ and thus $\beta$ from $\mathbf{n}$ in a circuit around the loop of interest as described in Sec. 4. The small measuring circuit makes this calculation non-local in a mild way, which is feasible for individual loops but less practical for datasets of many defects.

We also used an alternative, purely local calculation of $\boldsymbol{\Omega}$ that agrees well with the non-local calculation: The direction of $\boldsymbol{\Omega}$ is estimated to be parallel to:

$$\tilde{\boldsymbol{\Omega}} \equiv \nabla \times \mathbf{n} - \mathbf{n}(\mathbf{n} \cdot \nabla \times \mathbf{n}) = \mathbf{n} \times [(\mathbf{n} \cdot \nabla)\mathbf{n}] \tag{8}$$

whereever $\tilde{\boldsymbol{\Omega}}$ has nonzero magnitude. The assumption $\boldsymbol{\Omega} \parallel \tilde{\boldsymbol{\Omega}}$ can be justified since they are exactly parallel for a straight disclination along the $z$-axis with uniform $\boldsymbol{\Omega}$, and with an idealized director field profile given by

$$\mathbf{n} = \hat{p}_1 \cos(\phi/2) + \hat{p}_2 \sin(\phi/2) \tag{9}$$

which produces a $\pi$ winding of $\mathbf{n}$ about an arbitrary $\tilde{\boldsymbol{\Omega}}$. Here, $\phi$ is the angle in the $xy$ plane, and $\{\hat{p}_1, \hat{p}_2, \hat{\boldsymbol{\Omega}}\}$ is an orthonormal triad. We therefore expect $\tilde{\boldsymbol{\Omega}}$ to be a good approximation to the direction of $\boldsymbol{\Omega}$ except where the disclination is strongly curved, has a rapidly varying $\boldsymbol{\Omega}$, or has $\beta = \pi/2$ exactly (in which case $\tilde{\boldsymbol{\Omega}}$ has zero length). In contrast to the non-local measurement of $\boldsymbol{\Omega}$ on a circuit, $\tilde{\boldsymbol{\Omega}}$ is a local measurement depending only on neighboring voxels for finite difference calculations of first derivatives of $\mathbf{n}$.

A separate measure is needed to choose between $\tilde{\boldsymbol{\Omega}}$ and $-\tilde{\boldsymbol{\Omega}}$ for the direction of $\boldsymbol{\Omega}$, because Eq. 8 is odd in $\mathbf{n}$. We use the saddle-splay energy density expression:

$$\tilde{f}_{24} \equiv \nabla \cdot [(\mathbf{n} \cdot \nabla)\mathbf{n} - \mathbf{n}(\nabla \cdot \mathbf{n})]. \tag{10}$$

Applying this expression to the ideal disclination director field of Eq. 9 shows that $\tilde{f}_{24}$ is positive at a $+1/2$ wedge profile, negative at a $-1/2$ wedge profile, and zero at a twist profile (*18*). We find that $\tilde{f}_{24}$ alone is not a reliable measure of $\beta$ in the data, but that we can use the sign of $\tilde{f}_{24}$ to fix the sign of $\tilde{\boldsymbol{\Omega}}$ relative to the tangent. With this combination of $\tilde{\boldsymbol{\Omega}}$ and $\tilde{f}_{24}$, we obtain an estimate of the direction of $\boldsymbol{\Omega}$ with the correct sign of $\boldsymbol{\Omega} \cdot \mathbf{t}$, from which we obtain the angle $\beta$.

# 4 Data analysis

**Identifying defect loops in director fields:** We identify the defect set in a voxelated director field as the set of all voxels where the magnitude of the gradient in the director, $|\nabla \mathbf{n}| = \sqrt{\partial_i n_j \partial_i n_j}$, exceeds a threshold, typically 0.5–0.6. The defect set is divided into its connected subsets, each of which is a defect. To differentiate loops from other defect structures, we filter these connected subsets by their topological genus. The genus, $g$, which counts the number of holes, can be computed from the Euler characteristic, $\chi$, through the relation $\chi = 2 - 2g$. Because a connected subset is a collection of cubic voxels, we can treat the subset as a polyhedron. Then, we can compute the Euler characteristic as

$$\chi = V - E + F, \tag{11}$$

where $V$, $E$, and $F$ are the number of vertices, edges, and faces of the polyhedron, respectively.

A connected subset $\mathcal{D}$ is a collection of cubic voxels which has an easily counted number of vertices, edges, and faces, from which we calculate the Euler characteristic. To avoid spurious one-voxel holes, we inflate $\mathcal{D}$ by adding to it all of the nearest-neighbor voxels to each of the original voxels of $\mathcal{D}$ (without repetition). The computation of the Euler characteristic can be sped up using pre-computed information about possible voxel cluster motifs (*19, 20*). From the Euler characteristic we obtain the genus and identify loops as connected subsets with $g = 1$.

**Calculating a loop core:** Once a connected defect subset $\mathcal{D}$ has been identified as a loop, we obtain the director profiles and calculate $\mathbf{\Omega}$ at several points on the loop. We first identify a loop core, a one-voxel-thick thread through the disclination contour, to guarantee a nontrivial winding of the director field in a small circuit around the disclination. Our algorithm to compute the defect loop core $\mathcal{C}$ chooses a starting point $\mathbf{x}_{\text{start}}$ as the location of the smallest value of nematic order $S$ (simulation) or the largest value of $|\nabla \mathbf{n}|$ (experiment) in the connected subset. Then, successive points in the loop core are chosen as follows:

1. Construct a list $\{\mathbf{x}_i\}_{\text{elig.}}$ of eligible next points in $\mathcal{C}$, choosing from among the nearest-neighbor points to the most recently added point $\mathbf{x}_{\text{prev}}$ in $\mathcal{C}$, and selecting only those neighbors that are members of $\mathcal{D}$ but not already members of $\mathcal{C}$. Exclude any points that are in the list $\{\mathbf{x}_i\}_{\text{excl.}}$ of excluded points (defined below). Further specify that $\{\mathbf{x}_i\}_{\text{elig.}}$ cannot include any point that is a nearest neighbor of any point already in $\mathcal{C}$ besides $\mathbf{x}_{\text{prev}}$ and $\mathbf{x}_{\text{start}}$, unless this requirement leaves zero eligible next points in $\{\mathbf{x}_i\}_{\text{elig.}}$.

2. If $\mathbf{x}_{\text{prev}}$ is not a nearest neighbor of $\mathbf{x}_{\text{start}}$, but at least one of the eligible next points in $\{\mathbf{x}_i\}_{\text{elig.}}$ *is* a nearest neighbor of $\mathbf{x}_{\text{start}}$, then choose one of those points neighboring $\mathbf{x}_{\text{start}}$ as the next point $\mathbf{x}_{\text{next}}$ in $\mathcal{C}$ in order to close the loop. Otherwise, choose the next point $\mathbf{x}_{\text{next}}$ according to smallest $S$ or largest $|\nabla \mathbf{n}|$ among the options in $\{\mathbf{x}_i\}_{\text{elig.}}$. Append $\mathbf{x}_{\text{next}}$ to $\mathcal{C}$. If this operation is not possible because there are no eligible next points in $\{\mathbf{x}_i\}_{\text{elig.}}$, then we have made a wrong turn; we therefore remove $\mathbf{x}_{\text{prev}}$ from $\mathcal{C}$, place $\mathbf{x}_{\text{prev}}$ in the list of excluded points $\{\mathbf{x}_i\}_{\text{excl.}}$, and return to step 1 with $\mathbf{x}_{\text{prev}}$ taken from the previous iteration.

3. If $\mathbf{x}_{\text{next}}$ is a nearest neighbor of $\mathbf{x}_{\text{start}}$, and $\mathcal{C}$ contains at least ten points, then we consider the loop core closed. Otherwise, we repeat steps 1-2 until $\mathcal{C}$ is closed.

By following the path of largest $|\nabla \mathbf{n}|$ or smallest $S$, the loop core advances through the connected subset until it closes. In some instances steps 1-3 produce a curve that does not thread through the defect loop, but instead forms a small loop in one portion of the defect subset. In these cases we run the algorithm again with a different starting point, chosen by the next-largest $|\nabla \mathbf{n}|$ value or next-smallest $S$ value in $\mathcal{D}$. To obtain the loop normal $\mathbf{N}$, the loop core is fit to a circle, and the normal to the plane of the circle serves as $\mathbf{N}$. An arbitrary choice of sign for $\mathbf{N}$ determines the positive sense of rotation around the loop. The local tangent vector $\mathbf{t}$ to the disclination is determined by the vector difference between next and previous points in the loop core, with sign chosen by the positive sense of rotation.

**Calculating the rotation vector and reference director**  To calculate $\mathbf{\Omega}$ at a given point $P$ on the loop core, we first make a collection of directors $\{\mathbf{n}(\mathbf{x}_i)\}$ from points $\{\mathbf{x}_i\}$ that are near $P$ and nearly in the plane $\Theta$ transverse to $\mathbf{t}$. More specifically, $\{\mathbf{x}_i\}$ is the set of points with separation vector $\mathbf{d}_i$ from $P$ satisfying $d_{\min} \leq |\mathbf{d}_i| \leq d_{\max}$ and $|\arccos(\mathbf{d}_i \cdot \mathbf{t})| > \psi_{\min}$. We use $d_{\min} = 3$, $d_{\max} = 7$, and $\psi_{\min} \approx 0.61 \times (\pi/2)$, where lengths are given in units of the voxel spacing.

Next, for each point $\mathbf{x}_i$, we project $\mathbf{d}_i$ into the transverse plane $\Theta$, $\mathbf{d}_i^\perp = \mathbf{d}_i - \mathbf{t}(\mathbf{d}_i \cdot \mathbf{t})$, to obtain an angle $\phi_i$ in the plane relative to a fixed reference direction. The collection of directors $\{\mathbf{n}(\mathbf{x}_i)\}$ is then plotted on the unit sphere in order of $\phi_i$. A semi-circle is fitted to the path traced out by this ordered collection of directors, with a definite orientation given by the positive sense of rotation in $\Theta$, whose positive normal direction is aligned with $\mathbf{t}$. We thus compute $\mathbf{\Omega}$ as the positive normal direction to this fitted semi-circle on the unit sphere. We also take the reference director $\mathbf{n}_{\text{out}}$ to be the $\mathbf{n}(\mathbf{x}_i)$ at the point $\mathbf{x}_i$ in the collection for which $\mathbf{d}_i$ is most nearly outward from the center of the circle fitted to the loop. Any small component of this $\mathbf{n}(\mathbf{x}_i)$ along $\mathbf{\Omega}$ is projected out to obtain $\mathbf{n}_{\text{out}}$.

**Calculating disclination loop topology:**  We measure the topological index $\nu$ of a disclination loop as follows: At roughly even intervals around a loop separated by angle $\Delta\phi$, we calculate $\mathbf{\Omega}$ and $\mathbf{n}_{\text{out}}$ to find the frame $\mathbf{F}(\phi)$ at each step. We record the $SO(3)$ rotation operation $\mathbf{R}_{\phi,\phi+\Delta\phi}$ that takes $\mathbf{F}(\phi)$ into $\mathbf{F}(\phi+\Delta\phi)$ and use $\Delta\phi$ between $2\pi/12$ and $2\pi/8$. Expressing $\mathbf{R}_{\phi,\phi+\Delta\phi}$ as a rotation by angle $\alpha$ about an axis $\hat{a}$, we convert the rotation to a quaternion $q_{\phi,\phi+\Delta\phi} = \cos(\alpha/2)\mathbf{1} - \sin(\alpha/2)(\hat{a}\cdot\vec{\sigma})$, where $\vec{\sigma} = i\hat{x}+j\hat{y}+k\hat{z}$ (borrowing from the Pauli matrix language). The product $q_{0,2\pi} = \prod_{\phi=0}^{2\pi-\Delta\phi} q_{\phi,\phi+\Delta\phi}$ identifies the loop with one of the cases named in Sec. 3, as $q_{0,2\pi}$ must be one of the following: $1$ (unlinked and neutral), $-1$ (unlinked and charged), or a quaternion satisfying $q^2 = -1$ (linked).

**Figure S1**

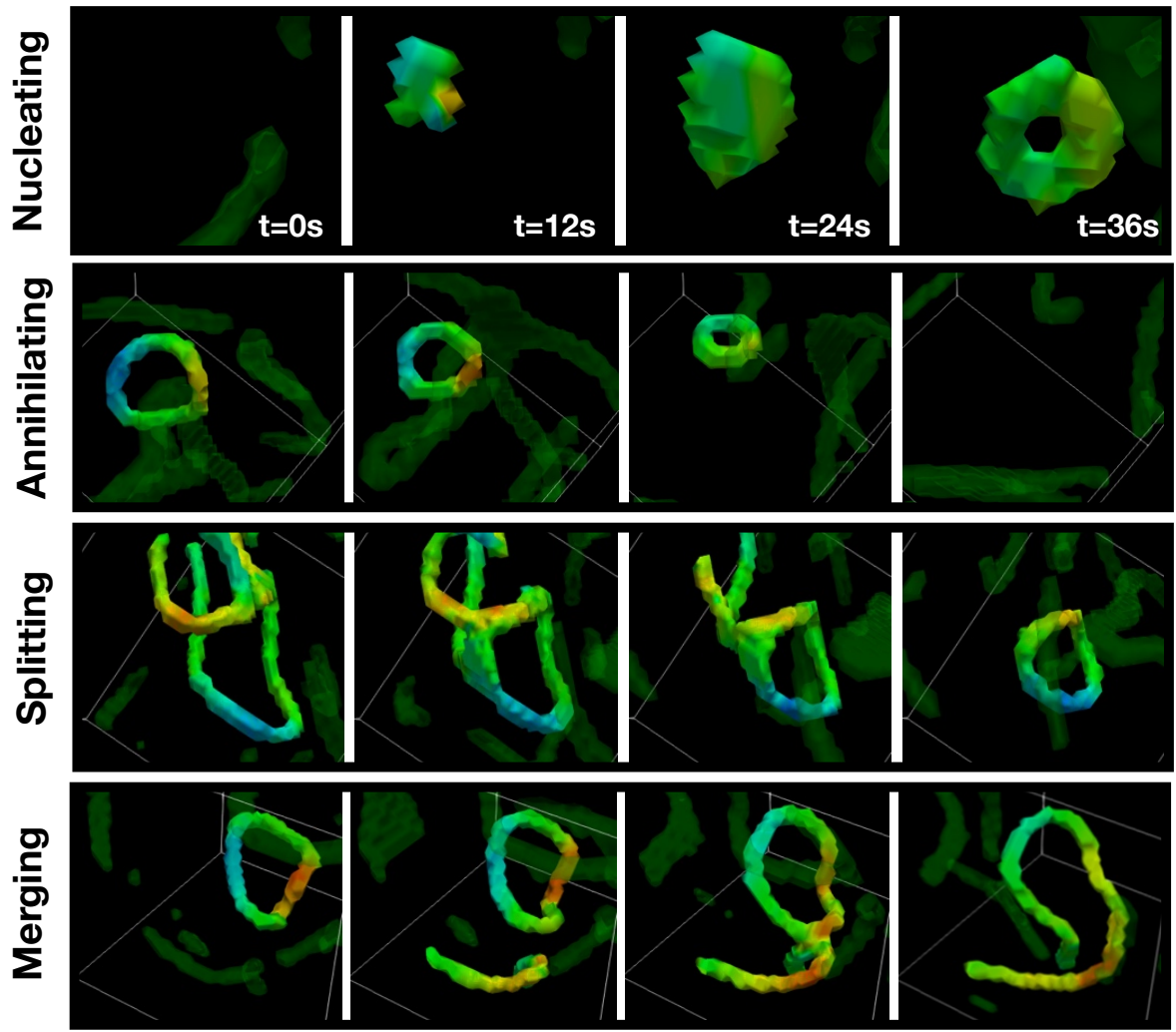

**Dynamics of disclination loops observed in hybrid lattice Boltzmann simulations.** (A) Loop nucleation from a defect-free region. (B) Loop self-annihilation leaves behind a defect-free nematic. (C) A disclination line self-intersects and splits, emitting a loop. (D) A disclination loop merges with a disclination line. Coloring of the disclinations indicates $\beta$ as defined in Fig. 3.

**Figure S2**

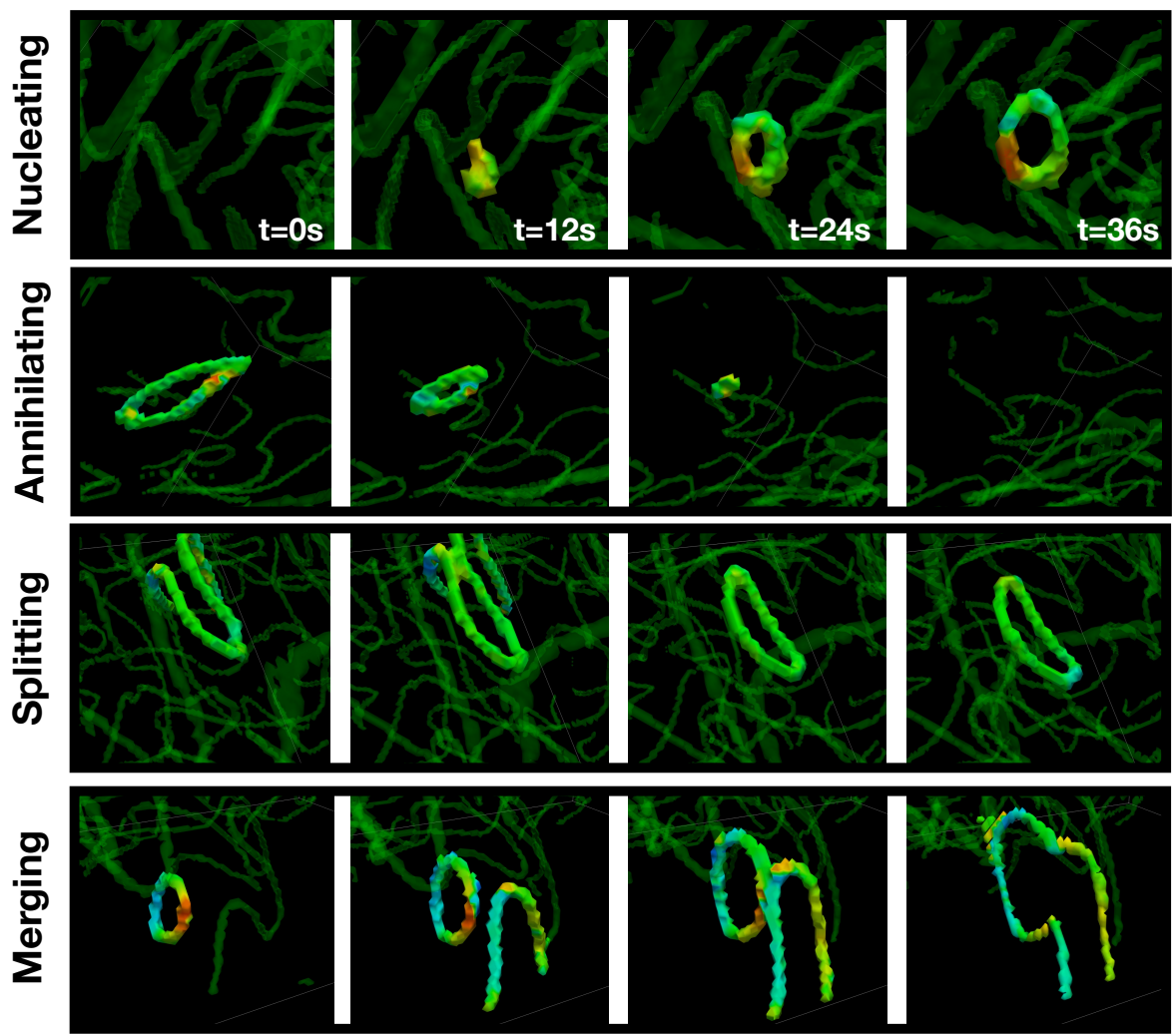

**Dynamics of disclination loops observed in finite difference simulations.** (A) Loop nucleation from a defect-free region. (B) Loop self-annihilation leaves behind a defect-free nematic. (C) A disclination line self-intersects and splits, emitting a loop. (D) A disclination loop merges with a disclination line. Coloring of the disclinations indicates $\beta$ as defined in Fig. 3.

**Figure S3**

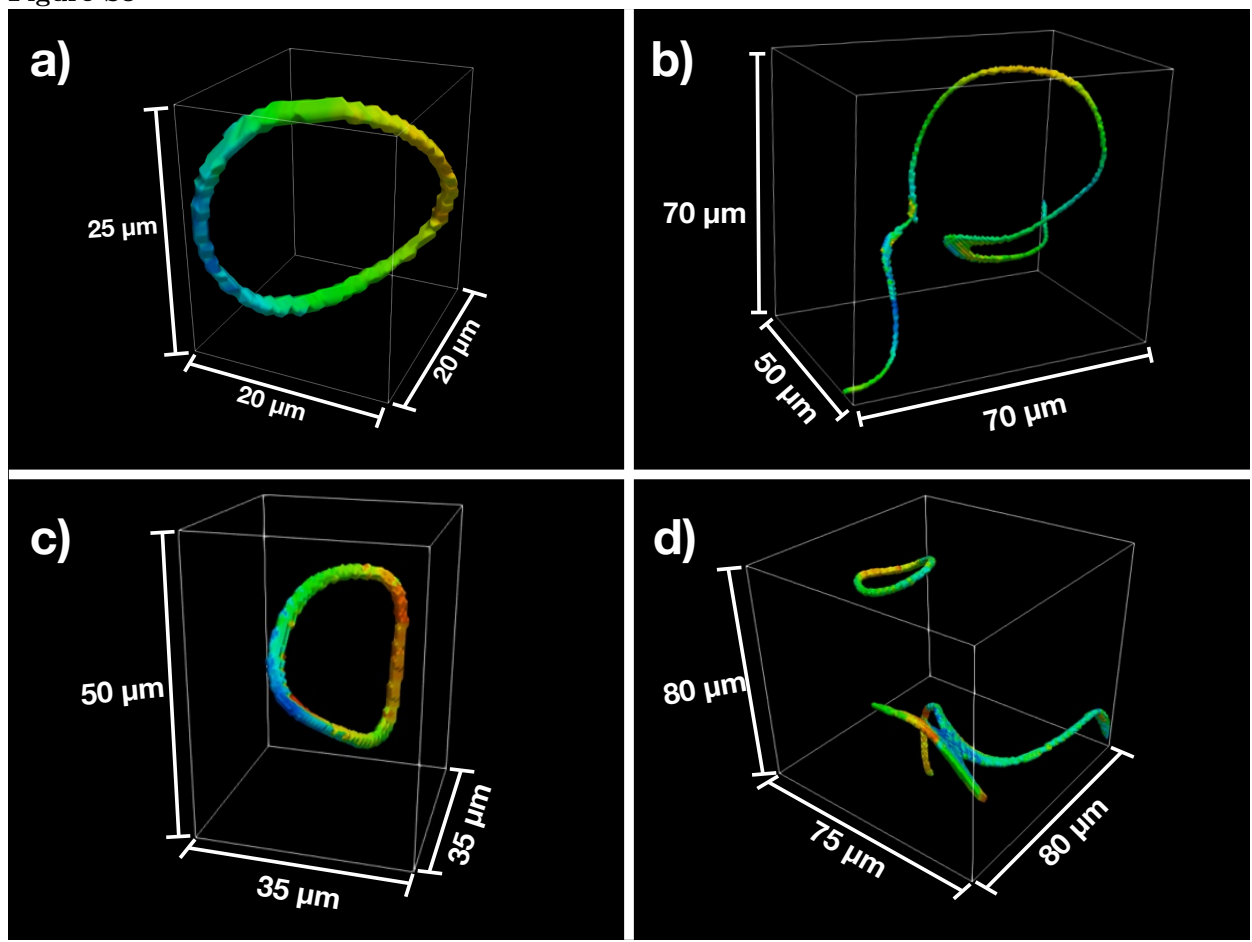

**Scale of the loops shown in Fig. 2.** Bounding boxes are shown to give the scale of the (A) nucleation event, (B) splitting event, (C) annihilation event and (D) merging event. Coloring of the disclinations indicates $\beta$ as defined in Fig. 3.

**Figure S4**

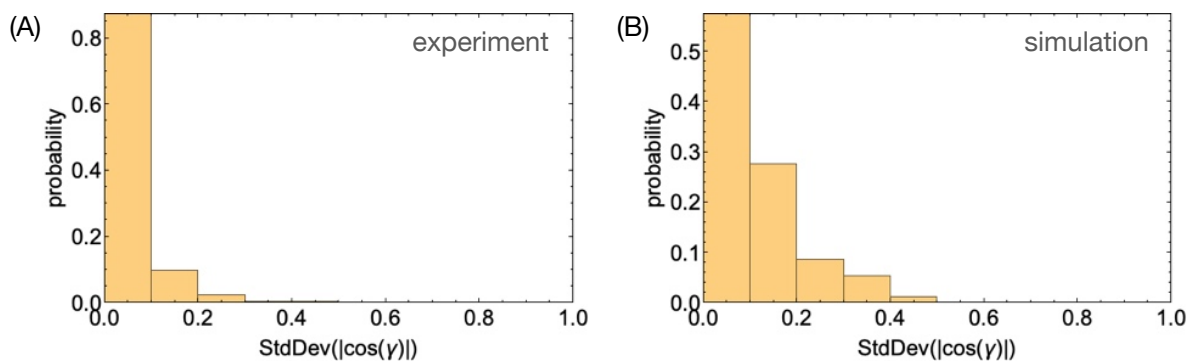

**Distributions of standard deviations of** $|\cos(\gamma)|$ for the (A) experimental (N=268) and (B) simulated (N=94) loops represented in Fig. 4D. Each standard deviation value pertains to one loop, with $\mathbf{\Omega}$ calculated at 8-12 approximately evenly spaced points along the loop contour.